LASER

# Fluorescence spectrum analysis using Fourier series modeling for Fluorescein solution in Ethanol

MAHASIN F. HADI (*)

SUMMARY. – We have measured the fluorescence spectrum for fluorescein solution in ethanol with concentration $1 \times 10^{-3}$ mol/liter at different temperatures from room temperature to freezing point of solvent, (T = 153, 183, 223, 253, and 303 K) using liquid nitrogen. Table curve 2D version 5.01 program has been used to determine the fitting curve and fitting equation for each fluorescence spectrum. Fourier series (3 × 2) was the most suitable fitting equation for all spectra. Theoretical fluorescence spectrum of fluorescein in ethanol at T = 183K was calculated and compared with experimental fluorescence spectrum at the same temperature. There is a good similarity between them.

## 1. Introduction

The photophysics and photochemistry of dyes in general are of considerable interest in the appreciation of various phenomena such as fluorescence, phosphorescence, long range and short range excitation energy transfer and other modes of quenching, as probes for liquid structure in mixed solvents and various relaxation processes in solution (1). In laser technology dyes are used for tuning, mode locking and Q-switching (2).

Dyes, either as solution or vapors, are the active medium in pulsed and continuous-wave dye lasers (3). The fluorescein dye is probably the most common fluorescent probe. Its very high molar absorptivity at the wavelength of the argon laser (488 nm), large fluorescent quantum yield and high photostability makes it very useful and sensitive fluorescent label (4).

Fluorescein in aqueous solution occurs in four forms: cationic, neutral, anionic and dianionic, making its absorption and fluorescence properties strongly PH dependent (5,6).

(*) Al-Mustansiriyah University, College of Science, Physics Dept., Baghdad-Iraq; e-mail: drmahasinf@yahoo.com

In the present work we measure the fluorescence spectra of fluorescein solution in ethanol at low temperature. We will find theoretical model for studying the effect of varying temperature on the fluorescence spectrum of fluorescein solution in ethanol and compare it with experimental work.

## 2. Experimental part

### 2.1 – *Materials*

Dye was used fluorescein which is crystalline powder (orange- red). This seems like a yellow- green solution when the dye is dissolved in very basic dilute solutions; and dissolved in hot alcohol, ether, and very dilute acids (7).

The chemical form of fluorescein is $C_{20}H_{12}O_{15}$ and its molecular weight is 332.30 gm/mol (8, 9). The chemical structure of fluorescein dye is shown in Fig. 1 (10, 11).

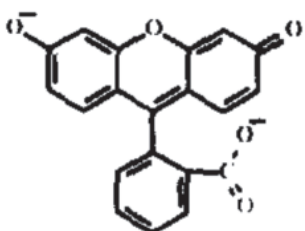

Fig. 1

The chemical structure of fluorescin dye (11)

The solvent used was ethanol. Its chemical form is $C_2H_5OH$, molecular weight 46.07 gm/mol, refractive index 1.364, and freezing point –114.1°C (8,9).

### 2.2 – *Preparation of dye solution*

To prepare the dye solution of a certain concentration one has to dissolve the proper amount of powdered dye in a certain volume of solvent used in accordance with this relationship:

$$m = \frac{C \times V \times M}{1000}$$

where

- $m$ = weight of the dye needed to obtain the desired concentration in unit (gm).
- $C$ = the concentration needed to prepared in unit mol/liter.
- $V$ = the volume of solvent in cm³ necessary to add to the dye.
- $M$ = the molecular weight of the dye used.

The concentration $1 \times 10^{-3}$ mol/liter for fluorescein solution in ethanol was used in this work.

### 2.3 – *Equipment used*

Fluorescence spectra emitted by fluorescein solution in ethanol were recorded by thee Jerrelash grating 82-410, 1meter Gzenny-Turner Spectrometer Spectrograph. We used the liquid nitrogen to obtain fluorescence spectrum at low temperatures.

Fitting curves for fluorescence spectrum were taken by the program Table Curve 2D version 5.01.

## 3. Results and discussion

### 3.1 – *Practical results*

The effects of temperature on the fluorescence spectrum of fluorescein solution in ethanol were studied for different temperatures from room temperature to the freezing point of solvent ($T = 153, 223, 253$ and $303$ K), as shown in Fig. 2. We found that the intensity of this fluorescence spectrum decreases with decreasing temperature. The fluorescence spectrum becomes narrower than it is at room temperature.

Thus, this fluorescence spectrum is subject to substantial changes with temperature changes. When temperature increases, the vibration and rotational energies increase as a result of thermal excitation lead to broading spectrum. But when temperature decreases, the broad band transfer to sharp band because of low thermal excitation, and emission lines arise from one or most of rotational levels, where the spectrum resulting from higher levels disappear due to decreasing number of atoms in these levels.

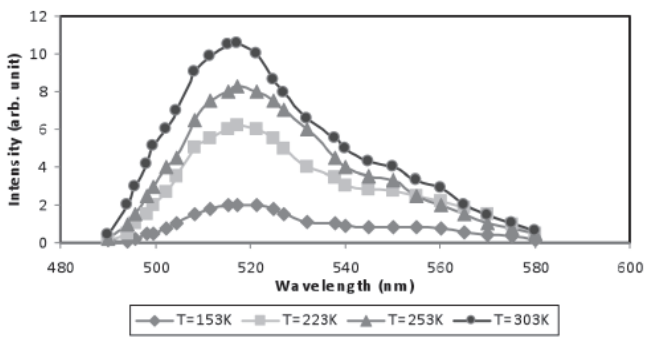

Fig. 2

Fluorescence spectrum of fluorescein solution in ethanol at different temperatures.

Lopez Arbeloa (12, 13) show that the increase in the temperature does not affect the shape of fluorescence spectrum of fluorescein. In dilute solutions, the small increase in temperature produces a little decrease in intensity. While, at high concentrations where the dimmer will be sensitive, the intensity of the fluorescence increases with temperature. And in very high concentrated solution, the increase in intensity with temperature will be very small and these changes are attributed to the increase of quenching arising from the monomer in dilute solutions. In the solution with high concentration, the increase of intensity of fluorescence spectrum caused by separation of dimmer. Whereas in most concentrated solutions, the stability of trimmer respect to dimmer explain the little increase in intensity of fluorescence spectrum with temperature

### 3.2 – Modeling results

For our practical results, we take the fitting curve for fluorescence spectrum of fluorescein in ethanol at temperature ($T = 153$, 223, 253, and 303 K). The resulted fitting curve for each case are shown in Figs. 3-a, b, c, and d, where we used "Fourier series polynomial $3 \times 2$"; given by

$$Y = a + b\cos\frac{\pi x}{L} + c\sin\frac{\pi x}{L} + d\cos\frac{2\pi x}{L} + e\sin\frac{2\pi x}{L} + f\cos\frac{3\pi x}{L} + g\sin\frac{3\pi x}{L}$$

where the values of these parameters $a$, $b$, $c$, $d$, $e$, $f$ and $g$ are shown in table (1) for each fluorescence spectrum.

**Table 1**
The parameters of fourier series equation for each temperature

| Parameter | T = 153K | T = 223K | T = 253K | T = 303K |
|---|---|---|---|---|
| $r^2$ | 0.9887655964 | 0.9964449279 | 0.9977214297 | 0.9968112955 |
| a | 4.720884126 | 12.03248502 | 14.71322672 | 16.13396709 |
| b | 0.292436399 | 0.699377217 | 0.938753298 | 0.504960462 |
| c | −6.45437877 | −15.4967383 | −18.8017568 | −19.5344177 |
| d | −4.48660452 | −11.4941435 | −14.0459413 | −15.3308608 |
| e | 0.027008741 | 0.414870470 | 1.161572929 | 2.472226161 |
| F | −0.35815844 | −0.94805488 | −1.08510976 | −0.78498411 |
| G | 1.642231403 | 4.432493318 | 5.219941360 | 6.204833349 |

Figures 4-10 describe the variation of each parameter with temperature, and we take fitting curve for each one, the fitting equation being illustrated above the curves. We used different fitting equations for these parameters.

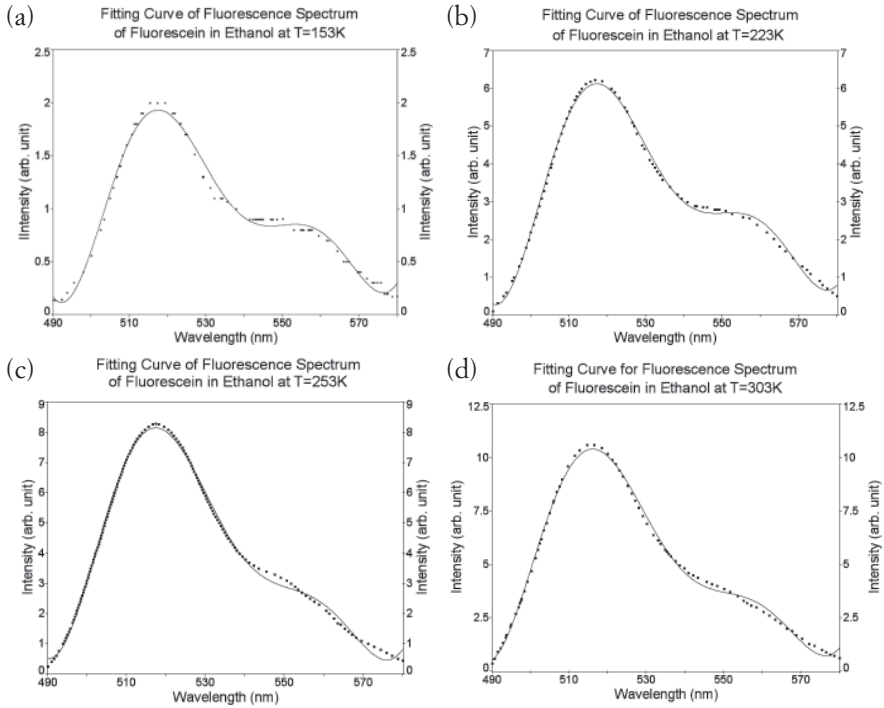

Fig. 3

Fitting curve for fluorescence spectrum of fluorescein solution in ethanol at different temperature (a) $T=153$K, (b) $T=223$K, (c) $T=253$K. (d) $T=303$K

fitting – equation
$\ln y = -5.88005262 + 0.195512058\,x - 0.02921478\,x \ln x$

fitting – equation
$y^{-1} = -58.522785 + 2.1681298 e^{-7} x^3 + 310.52722/\ln x$

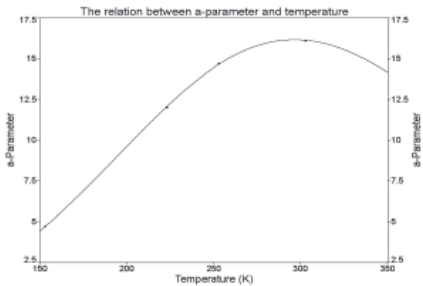

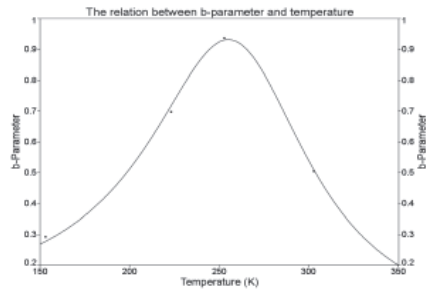

Fig. 4

The relation between a-parameter and temperature

Fig. 5

The relation between b-parameter and temperature

fitting – equation
$y^{-1} = -0.86085587 - 5.5179472 e^{-7} x^2 \ln x + 0.063129842 x^{0.5}$

fitting – equation
$y^{-1} = -2.7123023 - 5.1641077 e^{-7} x^2 \ln x + 0.51064328 \ln x$

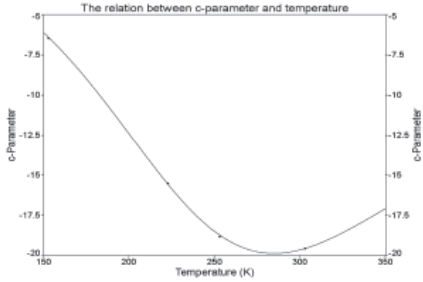

FIG. 6

The relation between c-parameter and temperature

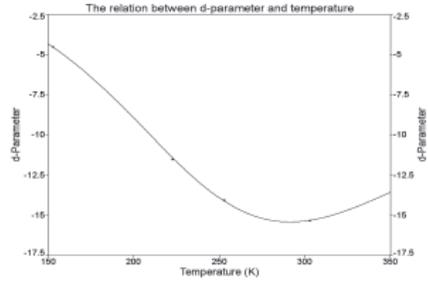

FIG. 7

The relation between d-parameter and temperature

fitting – equation
$y = 15.31519965 - 5476.38829 / x + 480014.7068 / x^2$

fitting – equation
$y^{-1} = -83.3747693 - 0.07257321 x + 18.21843232 \ln x$

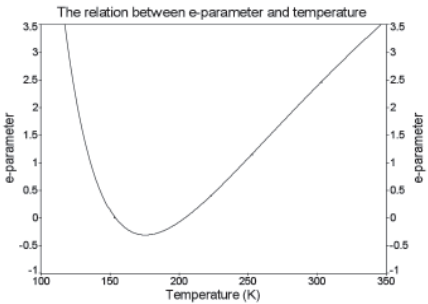

FIG. 8

The relation between e-parameter and temperature

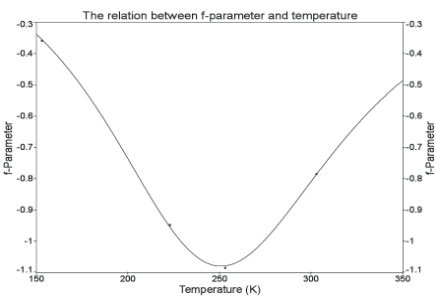

FIG. 9

The relation between f-parameter and temperature

fitting – equation
$y^{-1} = -2.072282 + 0.00339372171 x + 63.959332 \ln x / x$

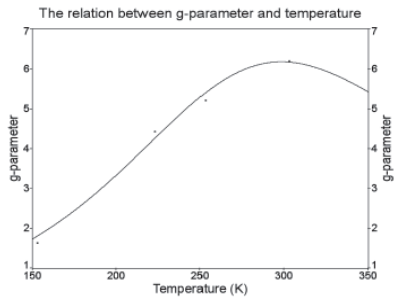

FIG. 10

The relation between g-parameter and temperature

After that, we calculated the final parameters from above fitting equations as test to find the theoretical equation for fluorescein solution in ethanol spectrum at T = 183 K, the estimated equation as follow

$$Y = 7.745827366 + 0.41425332\cos\left(\frac{\pi x}{90}\right) - 9.6976687\sin\left(\frac{\pi x}{90}\right)$$
$$-7.0320272\cos\left(\frac{2\pi x}{90}\right) - 0.27691569\sin\left(\frac{2\pi x}{90}\right)$$
$$-0.572410\cos\left(\frac{3\pi x}{90}\right) + 2.7063002\sin\left(\frac{3\pi x}{90}\right)$$

The estimated theoretical fluorescence spectrum as function of temperature is plotted and compared with experimental fluorescence spectrum at the same temperature in Fig. 11. There is a good similarity between them.

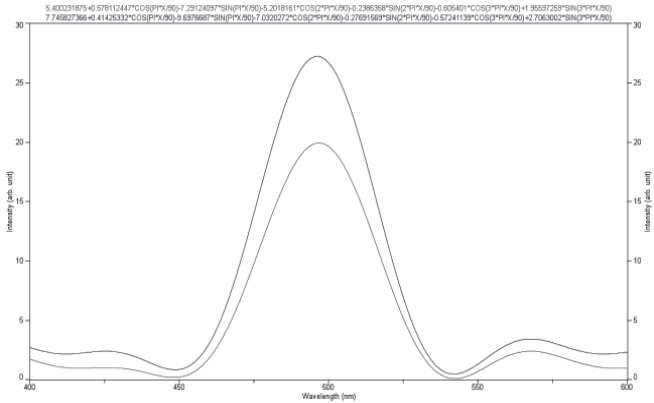

Fig. 11

Theoretical and experimental fluorescence spectrum of fluorescein solution in ethanol at T = 183 K
Upper curve = theoretical fluorescence spectrum
Lower curve = experimental fluorescence spectrum

## 4. Conclution

From spectroscopic properties of fluorescein solution in ethanol with concentration $1 \times 10^{-3}$ mol/liter, we predict theoretical model for the effect of temperature on the fluorescence spectrum. Fuoreir series $3 \times 2$ is the best fitting equation for our results, that is clear from the similarity of the theoretical and practical fluorescence spectrum at T = 183 K.